# The bubble-universe and a physico-chemical analogy of the cosmological constant.

C. Della Volpe
Applied Physical Chemistry Laboratory,
University of Trento, Via Mesiano 77, 38050 Trento (Italy).
Tel. +39-0461-282409  Fax +39-0461-281977
e-mail: claudio.dellavolpe@unitn.it
Trento, July 28th 2012

**Abstract.**

An analogy is developed between the bubble universe model and the Young-Laplace equation obtaining an analogic equation and commenting some numerical results produced by its application. The results, even if based on a simple analogy, appear quite intriguing and very close to the experimentally estimated or calculated values. Some conceptual and numerical limitations of such an analogic approach are also indicated.

# Introduction.

The term bubble-universe has been used in official cosmology papers (see as an authoritative example ref. [1], even if Guth prefers to use the term "pocket" universe for a very precise reason). It is however common to speak about the "bubble-universe", presenting in an evocative way some properties we attribute to the Universe, as we know it, even in the multiverse approach. It is certainly a pure analogy, but notwithstanding its limits, it may be useful to reflect on the models used in Physics.

However we can bring the analogy a step forward considering more in details a bubble-universe, as a quantum-vacuum phase, in contact with a false-vacuum phase, substantially following the model proposed by Guth[1].

What are the properties of such a system? What happens if two such phases meet? Applying the Young-Laplace equation to this system and opportunely substituting numerical values for its quantities some conclusion may be obtained; among others a surprisingly precise value of the Universe energy density (or alternatively of its cosmological constant).

Also the use of an analogic approach has a base in literature [2]; this approach may be justified considering that *"In studying gravity at the Planck scale, experiments cannot be performed because they are beyond our reach in energy."* [2]. Condensed matter analogs of classical, semiclassical and even quantum gravity effect may be found in literature [citations 2-7 of ref.2].

I am here neglecting all the (probably correct) objections on the peculiar nature of the interface we are considering, on the number of involved dimensions, etc etc.; I am simply reasoning by analogy.

# Two-phases multiverse.

In a common physical system two phases in contact may be or not at equilibrium; if they are at equilibrium, then between them an interfacial tension exists at their interface, which is positive; this quantity justifies that the interfacial surface area tends toward a minimum. On the contrary if the two phases are not reciprocally stable, if they tend to mix totally (if they mix only partially we reduce substantially to the first case) then a negative interfacial tension exists. The positive or negative sign of surface tension may be seen also as a matter of stability; a stable (or stationary) two phases system has a positive interfacial tension, an unstable system a negative one. In the first case the interfacial area increase has a free energy cost higher than the potential energy decrease of the system, while in the second case this cost may be paied and the disruption of the original system is the effect. The nice book by van Oss [3] on the acid-base theory of surfaces developes the thermodynamic theory of this approach, and even if there is no complete agreement on the operational definition of a negative interfacial tension and on its measurement, the concept may be considered as a well acquired one.

The unstable interface has the effect to push toward a complete and irreversible mixing of the two phases, producing a maximum area interface among the constituents of the two phases. Obviously there are some other possibilities, not simply a partial mixing, (which is however able to create a stable interface, modifying the composition of both phases) but also a situation in which the kinetics of mixing is not fast; in this way a temporary interface may exist; alternatively a gas bubble, formed in an oversaturated solution may spontaneously create on a wall defect, detach from it when/if the buoyancy force is sufficient and move toward the upper liquid surface, increasing its size while the external pressure decreases: this simple analogy is apparently very close to a simple-minded idea of our Universe and its expansion!

To apply the common equations of the surface thermodynamics, as the Young-Laplace equation[4], to the quantum-vacuum/false-vacuum interface we are doing the hypothesis that this two vacua may be considered as two thermodynamic phases; we have to choose between the two main cases, shown before, and find the way to express the interfacial tension of the two multiverse phases.

To do this we may reflect on a relatively simple property of the surface thermodynamic quantities. Those quantities are commonly the surface tension and the surface energy or surface eccess energy. They have a close (even not identical) physical meaning but the same dimensional ratio; it is a common procedure to consider them as interchangeable. Multiplying the numerator and the denumerator of the surface tension times a length we obtain a surface energy. In an analog way the pressure may be transformed in an energy density, and this is the case for the gas theory or the radiation pressure.

Finally, instead of multiplying a dimensional ratio for the same dimension we may consider the reverse operation: the simplification of a dimensional equation, dividing times the same dimension.

These transformations of the dimensional equations of the involved quantities may help us to find what we may call an equation-by-analogy.

## A cosmological by-analogy equation.

The Young-Laplace equation is the fundamental equation of the surface thermodynamics; it is a relationship between the pressure difference, $\Delta P$, and the curvature $k$, of each couple of phases in stable equilibrium contact through a curved interface:

$$\Delta P = \gamma k \qquad \text{(1a)}$$

$k$ is in fact the sum of the main curvatures, $k_1$ and $k_2$ along two perpendicular directions; this quantity may be substituted by the so called mean curvature H, equal to the mean of the two main curvatures; so the equation (1a) may be written as:

$$\Delta P = 2\gamma H \qquad \text{(1b)}$$

The surface tension, $\gamma$ has the role of a proportionality factor; if the curvature is positive (convex surface) and the surface tension is positive, then a positive pressure difference exists between the internal of a convex surface phase (e.g.:a drop or a bubble) and its external; using this equation all the capillaries effect may be analysed.

We may easily apply the previously invoked transformation/reduction method to the Young-Laplace equation.

We will have three modifications, one for each term of the equation to transform it, by analogy, in a relation between "cosmological" quantities.

The pressure, or better its difference, may be transformed in an energy density difference, between the two phases in contact, quantum vacuum and the false vacuum, so we may substitute to it an energy density or an energy density difference $\rho_{E/V}$.

In the case of surface tension we may simply apply the simplification of the dimensional ratio, reducing the surface tension to a mass/time$^2$ ratio ([M]/[T]$^2$; we are simply considering that force is not necessarily a primitive quantity.

Such a simplification has no meaning in a common laboratory practice, but it may acquire a sense in the multiverse system; instead of considering a force per unit length we may consider the ratio of mass to time at second power, with the time approximately substituted by the age of the universe. Thus we have a term $M/t^2$ instead of $\gamma$.

Finally the curvature, which has the dimensions of a reciprocal length, may be susbstituted with the square root of its second power; the dimensions of a reciprocal second power length are the same of the cosmological constant (CC), which may be seen as the "analog" of a curvature, thus instead of $k$ we have a $\sqrt{\Lambda}$.

This transformation should be analysed more in detail. A quantity with the same dimensions of the CC is the Gauss curvature, K. Now substituting to H the quantity $\sqrt{\Lambda}$ we are in fact supporting that H$^2$=$\Lambda$ or in other terms H$^2$=K; this is true ONLY in umbilical points, because in general H$^2\geq$K; so conclusively, given that ONLY on a spherical surface all the points are umbilical, doing this hypothesis we are also doing the hypothesis that the universe surface is spherical, which is not a commonly accepted idea; in the paper by Guth cited at the beginning Guth uses the term "pocket universe" exactly to criticize the idea that the universe has a spherical shape; so the previous position is not in agreement with Guth hypothesis.

Considering all the dimensional and by-analogy transformation we have for the interface between a bubble-universe expanding in a Multiverse phase and in contact with it:

$$\rho_{E/V} = \frac{M}{t^2}\sqrt{\Lambda} \qquad (2)$$

## Some numerical tests.

We may perform some simple numerical test on the newly introduced quantities and on the equation (2).

If the surface tension of the bubble-universe may be considered as the ratio between its mass and its age at the second power, then its surface tension is by far higher than all the surface tensions we know for common materials. The water surface tension in MKS units (J/m$^2$) is of the order of 0.07; the highest experimental values of the surface tension for ceramic materials are of the order of 1-3 in MKS units (I am neglecting here the *vexata quaestio* of the difference between the surface tension in a liquid and a solid); in the same units we should have for the quantum/false vacua interface 10$^{17}$. The value of surface tension is generally related to the level of interacting forces among the matter particles or the particles constituting the phase; the ratio of the ceramic solids and water surface tensions is more or less depending on the ratio between the hydrogen-bonds and the ionic bonds energies. But what about a 10$^{17}$ surface tension? Even the gluonic forces are generally considered only a couple of orders higher than the electromagnetic interactions. This huge cohesion force should be considered as depending on the cohesion of the quantum foam [5] or of the 4-dimension simplexes constituting the space time [6], reasonably due to the extremely high forces acting at the Planck scale. Are there other unknown and stronger forces? A conclusion to be drawn is that, IF the cohesive forces of quantum vacuum are so high, then there is a good reason why it does not mix with the false vacuum.

It is not however a surprise that similar or even higher values of "surface tension" are evaluated for cases where elementary particles or quarks are considered.

As an example in the so-called "liquid drop" model, which is one of the traditional approaches to the nucleus energy the surface term have the value of 17-18 MeV for each nucleon; this quantity, considering the effective area of a nucleon, is very close to the value estimated for the universe.

In fact 1 eV=$1.6 \times 10^{-13}$ J and the nucleon radius is $10^{-15}$ m (1fm) so that its area is roughly 4 x $10^{-30}$ $m^2$ (4$fm^2$); as a consequence the surface term per nucleon is 0.2 x $10^{17}$ $J/m^2$, only slightly lower (consider that 1MeV/$fm^2 \sim 10^{17}$ $J/m^2$).

Another interesting case is the value of quark surface tension in a nucleon; also in this case the estimated or calculated values are very close: between 7 and 30 MeV/$fm^2$ [7]. Finally for the so called "strange stars" a value of 100-200 MeV/$fm^2$ has been estimated [8].

Conclusively the surface tension esitimated by analogy for the quantum/false vacua is of the same order of that estimated for elementary particles or even lower.

A good question may be: is it acceptable that a lower surface tension container (the quantum/false vacua interface) has in its internal part higher surface tension elements (common matter, strange stars, nucleons)? The answer may be yes, because this is exactly what happens in more common cases: lower surface tension materials are "exposed" to environment to reduce the interfacial tension and stabilize the system. There is even a similar proposed general rule based on this idea for biological tissues[9]. Fascinating hypothesis: cells tissue and universe structure driven by the same simple mechanism (or probably I am superposing local matter structure to exotic situations)!

A second point is about the temperature coefficient. We know that temperature coefficient of common matter surface tension is negative, i.e. a cold material has a surface tension higher than the same material at higher temperatures. This depends on the fact that the temperature derivative of the interfacial free energy is the negative of surface entropy; given that in common systems surface entropy is a positive quantity and that the temperature increase corresponds to an increase of surface entropy, surface tension decreases with increasing temperature.

The quantity $M/t^2$ has a different behaviour; in fact we know that the initial temperature of the Universe was very high and consequently we may expect that the surface tension was very low; but this is not the case for the $M/t^2$ quantity; it decreases with increasing time and thus with decreasing temperature. It corresponds to an initial infinite or however very high surface tension (to eliminate the infinity we could consider that the time began at least an attosecond before the big-bang explosion). Does this mean that the interfacial entropy of quantum-vacuum/false-vacuum has different properties than common matter? The initial Universe had also a very high value of density and of surface curvature; surface curvature at Planck size is probably very high yet now; we know that both quantities may strongly modify the surface tension of a system.

An alternative possible solution is that the surface entropy of bubble-Universe interface may have a behaviour similar to the black-hole entropy: it may increase with the area; if so, given the Universe espansion, it decreases with the temperature.

In a recent paper a similar idea for the trend of nuclear matter "surface tension" is proposed by Berger[10] for strange-quark matter of early universe. The high values estimated by Berger for the strange stars are also in agreement with the previous idea that the Universe at its beginning had a higher interfacial tension.

(The final answer is that, after all, an analogy cannot be stripped too much.)

We may now consider the cosmological constant, which appears in some way related to the curvature of the Universe; given the curvature is the square root of the CC, then its value is of the order of $10^{-26}m^{-1}$; this value corresponds to a radius of the order of $10^{26}m$, which, if interpreted as the radius of a spherical Universe, is of the correct order, between 10 and 100 billions of lightyear. Not so bad.

We may now arrive to the most intriguing aspect of equation (2); the quantities present in the equation can be evaluated separately; thus we may check the validity of the relation simply substituting the commonly accepted values of the different quantities and consider the result. Many cosmological theories find a very strong disagreement of the considered paremeters, e.g. it is common to speak about a paradox of the cosmological constant, for a disagreement of 120 orders of degree in some cases; what is the situation for our analogy?

The result is surprising.

We may found commonly accepted values of the different quantities present in equation (2) in the literature:

- M= $10^{53}$ kg (considering an universe made of $10^{11}$ galaxies each one made of $10^{11}$ stars with the mass of Sun, this is an accepted evaluation of the Universe mass[11a-b-c])

- t= 13.7 billion of years or $4.3*10^{17}$s; this datum is known very precisely[12]; $t^2=1.86*10^{35}$ $s^2$, supposing that the Universe age coincides with the time length (what may be not completely correct; does the beginning of the time coincides with the age of the Universe?);

$-\Lambda \cong 10^{-52}$ $m^{-2}$, a value estimated by the High-Z Supernova Team and the Supernova Cosmological Project [13].

Consequently the density $\rho$ becomes $\cong 5x10^{-9}$ J/m$^3$ and considering the Einstein equivalence between mass and energy, it becomes $\rho \cong 5x10^{-26}$ kg/m$^3$ or $\cong 5x10^{-23}$ g/m$^3$ or $\cong 5x10^{-29}$ g/cm$^3$ ; as a maximum, introducing the factor 2 of the equation (1b) $\cong 10^{-28}$ g/cm$^3$ while its experimental estimate (based on a spherical-like Universe with a diameter of about 92 billion year light[14]) is $10^{-29}$ g/cm$^3$[15]: the agreement is surprisingly good for a simple analogy!

A second positive result of the by-analogy equation is that it foresees that the Universe density decreases with time and consequently it was very high (even infinite) at the initial time; it is so in agreement with the BigBang model.

Given the ratio between mass and 2nd power of time is positive, a positive surface tension corresponds to a non-mixing multiverse, an Universe whose quantum-vacuum does not interact easily with the false vacuum; two mixing phases certainly wet well reciprocally (water and ethyl alcohol or water and ethylene glycole) ; on the contrary two non mixing phases do not interact very well; but in some cases one of them may wet the other (a good example are water and benzene, benzene (at least initially) wets water, but the two are immiscible).

So from this analogy the Universe espansion is not seen as the mixing between the two phases, but as the expansion of a phase into another one, exactly as a bubble of gas expanding in a liquid: after all the bubble-Universe analogy appears a self-supporting one. However we cannot exclude that the process is so fast that the two phases do not mix initially, but they may mix after a certain time or that they are non mixing but one may "wet" the other.

By analogy one can consider what happens mixing water and a hydrocarbon; the reciprocal solubility is very very low, but water is by far less soluble in hydrocarbon with respect to the hydrocarbon solubility in water. Similarly quantum vacuum may be considered as non-soluble in false vacuum, but false vacuum may be relatively more soluble in quantum vacuum. Using the analogy of an air bubble in water the reciprocal solubility is very low and a strong interfacial tension exist; but however the solubility of water (vapour) in air is of the order of 2-3%, far higher than the air solubility in liquid water.

A more complex case could be that quantum vacuum and false vacuum may "react" each other, as in the case of reactive wetting[16] , where the interfacial energy and the reaction energy sum up. Their reaction energy could eventually help to destroy the coherence of the highly cohesive quantum-vacuum Universe or even to stimulate the decay of false vacuum.

## Conclusions.

An analogy has been developed between the bubble universe model and the Young-Laplace equation obtaining an equation and commenting some numerical results produced by its application. The results, even if based on a simple analogy, appear in qualitative agreement with  experimentally estimated or calculated values of cosmological quantities, as the CC or the interfacial tension of quantum/false vacua and its temperature dependence. However the analogic procedure has many limitations and some conceptual and numerical limitations of such an analogic approach have been also indicated.

# References.

[1] A. Guth, arXiv:hep-th/0702178v1

[2] Joseph Samuel and Supurna Sinha "Surface Tension and the Cosmological Constant", PRL 97, 161302 (2006)

[3] C. van Oss, Interfacial Forces in Aqueous Media, CRC 1994

[4] T. Young, Phil. Trans., 1805, p. 65; P.S. de Laplace, Mécanique céleste, Supplement to the tenth edition, pub. in 1806

[5] M. Brooks Lewes, New Scientist, vol 28 (2191), p. 28 1999

[6] J. Ambjørn, J. Jurkiewicz and R. Loll  The Universe from Scratch
 arXiv:hep-th/0509010v3

[7] Marcus B. Pinto, Volker Koch, andJørgen Randrup  The Surface Tension of Quark Matter in a Geometrical Approach  in arXiv:1207.5186v1 [hep-ph] 21 Jul 2012

[8] R. Sharma∗ & S. D. Maharaj J. Astrophys. Astr. (2007) 28, 133–138
On Surface Tension for Compact Stars

[9] Ramsey A. Foty, Cathie M. Pfleger, Gabor Forgacs and Malcolm S. Steinberg
Surface tensions of embryonic tissues predict their mutual envelopment
behavior Development 122, 1611-1620 (1996)

[10] Micheal S. Berger Surface tension of strange-quark matter in the early Universe PHYSICAL REVIEW 40, 6 15  1989

[11] a) McPherson, Kristine (2006). "Mass of the Universe (http://hypertextbook.com/facts/2006/KristineMcPherson.shtml)". The Physics Fact book.

b) "On the expansion of the universe (http://www.grc.nasa.gov/WWW/K-12/Numbers/Math/documents/ON_the_EXPANSION_of_the_UNIVERSE.pdf)" (PDF). NASA Glenn Research Centre.

c) Helge Kragh (1999-02-22). Cosmology and Controversy: The Historical Development of Two Theories of the Universe. Princeton University Press, 212. ISBN 0-691-005.20.

[12] "Five-Year Wilkinson Microwave Anisotropy Probe (WMAP) Observations: Data Processing, Sky Maps, and Basic Results".

[13] Moshe Carmeli, Tanya Kuzmenko, Value of the Cosmological Constant: Theory versus Experiment arXiv:astro-ph/0102033 v2 4 Feb 2001 and citations therein

[14] Lineweaver, Charles; Tamara M. Davis (2005). "Misconceptions about the Big Bang (http://www.sciam.com/article.cfm?articleID=0009F0CA-C523-1213-8523pageNumber=5&catID=2)". Scientific American.

[15] http://hypertextbook.com/facts/2000/ChristinaCheng.shtml

[16] P.-G. de Gennes 1997 Europhys. Lett. 39 407-412

Note: it is possible also to consider a different analogy: the false vacuum as a high energy phase (a gas phase) decaying in a lower energy phase (a liquid phase), the quantum-vacuum; the condensation extends the low energy phase; this extension is the Universe expansion; a positive interfacial tension exists between the two phases; this further analogy could be considered elsewhere.